# opinion dynamics driven by
# leaders, media, viruses and worms

Çağlar Tuncay
Department of Physics, Middle East Technical University
06531 Ankara, Turkey
caglart@metu.edu.tr

**Abstract**

A model on the effects of leader, media, viruses, and worms and other agents on the opinion of individuals is developed and utilized to simulate the formation of consensus in society and price in market via excess between supply and demand. Effects of some time varying drives, (harmonic and hyperbolic) are also investigated.

## 1. Introduction

Within the existing literature about formation of consensus in a society in terms of opinion dynamics, several models and approaches were developed and utilized. Recently, we applied opinion dynamics to econophysics. (See, [1] for Sznajd model; [2] for a review; [3-6] for voters model; [7] for our application to market.) In the present literature focuse will be on the effects of leaders and media, where some more complicated agents, such as viruses and worms are also considered. Model is displayed in the following section. Applications are performed in the next one. Last section is devoted to conclusion.

## 2. Model

We site 10,000 people in one-dimensional matrix as our society, where any individual (I) is characterized by real numbers; two of which are her psychological mood a(I), and bond strength b(I, J) with the member (J). The opinion action on (surrounding opinion about, or the information gathered by) her is A(I); so, resulting opinion of her, i.e., her reaction will be O(I)=a(I)A(I).

The individuality term (a(I)) involves mainly the openness of a member to be effected by opinions of else. O(I) will be negative for a positive A(I) (and vice versa), if she is antithetical (of opposite, reverse character), i.e., a(I) < 0. The connection b(I, J), may change from duals of person (door neighbors, friends, relatives, etc.) to other ones, and distribution may depend on the quality of society.

Leaders are some inhabitants of the society, and they have fixed opinion (influence strength, ($O_{leader}$)). They are not effected by opinions of other members of the society; but, they effect them through personal interactions. Media is naturally not inhabitant of our one-dimensional society matrix, yet she broadcasts to the society a fixed opinion (influence strength of media), which is added to opinion (O(I)) of each individual at the end of an interaction tour. Therefore media effect an individual's opinion in two ways; directly (in terms of addition of the opinion broadcasted by them to the individual's resulting opinion O(I)), and via the interactions with members who are already effected by media.

*Individuality and connections*

In civilized societies b is expected to be small (weak connection) and almost uniform when compared with devoloping ones. In a scientific society many colleagues may effect the opinion of the individual under consideration; number of contacts per person will be big, and b is expected to be close to unity and almost uniform. In a market society, number of bonds per person will be few. Because, people do not talk extensively on their financial matters in general they consider their financial issues as private and secret. Yet, they share their financial privacies with some officials (dealers, brokers, etc.) and with some friends an relatives. So, b may be dispersed between zero and unity. Within a scientific society there may not exist any antithetical member, but in exchanges there exist many, especially within dealers and medium-big portfolio players. Two other reasons, besides many possible ones to support the present approach about the contribution of antithetical people may be underlined as: Anybody in any society may react oppositely from time to time. And, in a market the whole amount of sellings are bought by antithetical people during recession (and, vice versa).

We may assign four citizens (I-2, I-1, I+1, I+2) to be the most strongly bounded ones to any member (I), with bondstrength of unity. And, another four (I-4, I-3, I+3, I+4)to be relatively weakly bounded with a bondstrength of half a unity. We then compute the opinion formed around any individual (which we call the action, A(I)) as a weighted average of the opinion of the connected people over bonds with the pronounced bondstregths. So, $A(I) = (\sum O(\alpha)*1.0)/4 + (\sum O(\beta)*0.5)/4$, where the summations are performed over some $\alpha$, and $\beta$ which designates the primary (i.e. those with a bondstrength of unity) and secondary (i.e. those with a bondstrength of half a unity) strong connections of (I), respectively. Other connections are ignored.

*Initial distributions of opinions*

Several initial distributions may be performed to satisfy the needs of any given stituation [8]. In Figure 1. we exemplify a case, where initial opinions of arbitrarily chosen 961 people (out of 10,000 citizens) are randomly distributed between -1 and 1. An interesting feature is that, the envelope for the opinions decrease exponentially with number of interaction tours, (which may be taken as time (t)), i.e., $O(I) \propto \exp(\sigma t)$, with $\sigma<0$.

*Memory effect*

The action A(I) involves the effect of interacting people. A random distribution is utilized for the individuality term (a(I)), and the multiplication term a(I)A(I) may be taken as the resulting opinion formed at a given time. Yet, any citizen of the society may have a memory (non zero intelligence, self-confidence) as another characteristic feature besides these of her involved within a(I). We incorporate the effect of individual's memory, i.e. her previous opinion by incorporating it in the formation of the resulting opinion: $O(I)_{new} = O(I)_{previous} + a(I)A(I)$. In this case, envelopes of the opinions increase exponentially with time (t), i.e., $O(I) \propto \exp(\sigma t)$, with $0<\sigma$. (See, Figure 2.) In fact, memory effect may differ from person to person within any society, and the term $O(I)_{previous}$ within the previous summation should be multiplied by the pronounced distribution function. In our computations we multiply $O(I)_{previous}$ by either zero for no memory (zero intelligence) case, or unity for full memory (non zero intelligence, selfconfidence) case.

*Constraints on opinions*

In reality, opinions do not grow without limits. Mostly they turn into decissions, such as accepting an idea or rejecting it, voting for a candidate (person or a party) or not voting, buying in a market or not buying, selling in a market or not selling, etc. We may designate the positive (negative) decission by an opinion having the value +1 (-1), and stop the growths of opinions beyond it (them). Furthermore, we may apply threshold for decissions, as in Figures 3 a. and b., and Figures 4 a. and b., where the thresholds are ±0.999, and ±0.8, respectively. If an opinion is greater than the positive (less than the negative) threshold, we take it as (minus) unity. As thresholds decrease in absolute value, evolution saturates more quickly, since decissions are made more easily, opinions accumulate at -1 and +1 terminals more quickly.

It is worth to note that, within the further computations we reset parameters to their initial values at the end of each 100 tours, i.e. what is usually called iterations in literature (each site treated once in each tour). We call the time domain between each reset, a day. And five days make our week. Initial opinions will be randomly distributed between -1, and +1 (as in Fig. 3. a. and Fig. 4. a.; at t=0, and at t=100). Please notify that, in our market applications all the members with opinions $0.999 < O(I)$, i.e. the acceptors with opinion equals to +1, will be considered as demanders (buyers), and these with opinions $O(I)<-0.999$, i.e., rejectors with opinion equals to -1, will be considered as suppliers (sellers). Excess will be computed as the difference between demand and supply, explicitly; excess = demand – supply.

Now we may consider some complicated agents as leaders and mass media.

*Leaders and media*

What we meant by a leader was an inhabitant of the society who has a fixed opinion, i.e., influence strength ($O_{leader}$). We run the the leader effect with various influence strengths from zero to unity and we vary the number of leaders as 1, 10, 100 and 1,000. It was our previous[7] observation that, a single leader, do not contribute much to the opinion evolution in a crowded society with moderate influence strength and with small number of interaction tours. Effect of single-site leadership fades down, when the tours are extended. In order to establish a consensus within a society and to speculate on a market, one needs many leaders.

A short-cut is the mass media, where the opinion advertised by the mass media (influence strength of media) is added to these of individuals' formed at each interaction tour. It is known that, "The larger the lattice is the smaller is the amount of advertising needed to conceive the whole market."[9]. We found that media with influencial strength ranging between 0.1 and 0.15 is sufficient to create amply effects within the present society. Because, media influences an individual in two ways; directly (in terms of addition of media effect to the individual's resulting opinion of O(I)=a(I)A(I)), and via the interactions with members who are already effected by media.

*Viruses and worms*

In some cases, like the legal social organizations as political parties (underground ones like mafia, etc.) people may join to the leader with a=1, and b=1. And the number of joined people, as repeaters of the leader message, may increase in time. More new people may join as the number of joined people increases. This stituation, i.e. grow in the population of joined people is like the spread of living cells by copying themselves.

On the other hand, viruses may infect the computers, which are utilized to process in market, and viruses may spread throughout the net.

In all these events, new agents emerge with a=1, and b=1. Please imagine that, we start with a mostly empty chain, and slowly fill it with new agents. If we could expand the fully occupied chain from length N to length N+1 one a new agent is born. Suppose that N is the number of joined agents at any time, and $\Delta N$ is the increament of them in the coming time interval of $\Delta t$. Then we may define the rate of change $\sigma$ in N as $\sigma = \Delta N/\Delta t$, where the unit of dimension for $\sigma$ is (number of new agents)/(tour of opinion interactions). To be more precise, the dimension (tour of opinion interactions) maybe considered as (time), or (day). It is known that, $\sigma$ is proportional to N at a given time for biological viruses, and (after some calculus) we may state that $N = N_0 \exp(\delta t)$, where $\delta = \sigma/N$=constant.

On the other hand, $\sigma$ may be (independent of N, and) constant in time, as another commonly met variation. Some of the corresponding stituations in reality are: Conquering countries by an army, spreading of a message delivered by a very convincing leader such as a prophet who is traveling, spreading of a news through some very selected and reliable friends, buying of a big trading company for her clients in a market, and an internet worm traveling from computer to computer which are connected to market on the aim of processing there, etc.

## 3. Applications

We apply our model in market. The reasons for the present test are: First of all, market results are easy to check with reality. They are more familiar with respect to many other opinion dynamics applications, and they are more easily comprehendible. Any il-logical (wrong) approach in the formalism may be reflected throughout the calculations as an abnormal (contradicting) result, which may be easily detected. Many basic and important concepts and terms of the model becomes concrete (solid) in market jargon. For example, "acceptors (rejectors) of an idea" becomes "buyers (sellers)", etc. We may neatly see the effect of changing opinions in market, since we may count the number of buyers, sellers, waiters, and their total effect on price, which is an easily observable and measurable quantity. And finally, beyond all these, market is (by itself) a society, where opinion dynamics is the basic mechanism for processes. Moreover, we have many real markets in our life, with which we may compare various results of our simulations.

*Initial distribution of opinions*

Throughout the present work, we apply a random initial distribution between -1 and +1, and utilize threshold values of –0.999 and +0.999 to trap the rejectors of a proposed idea (sellers, anode side), and acceptors (buyers, cathode side), respectively. We count the accumulation of trapped opinions at anode and cathode sides, as displayed in Figure 5. a. and b. together with the corresponding excess (upmost curves in Figs. 5. b.-7. b.).

*Evolution of opinions in market society*

What characterizes a market society is that, her members come together to gain money. Some gain, while some lose. Any of them wants to know the opinion (decission, if possible) of all the others. Because, the price is shaped by the behaviors of all the people in market. Yet, they may behave very differently to the surrounding opinion A(I). Individuality a(I) involves diversity, and O(I) is expected to display dispersion in value. Connection b is not much extensive.

The society, with the pronounced parameters, come together every market morning with some opinions in mind, that we called (and reset to) the initial distribution of opinions of the market day. And accordingly, those with opinions above the trapping threshold for buyers are counted for demand, and those below the trapping threshold for sellers are counted for supply.

Within the absence of any leadership and mass media effects, the opinion dynamics takes place and survive inbetween the members which are very similar to each other. We have characteristic horizontal- (and occasional non-horizontal-) trends in excess, since randomness is dominant. Contribution of antithetical people averages to zero within medium- or long-run. And in short-run, antithetical contribution within moderate amplitudes may cause to stiff rises and falls, which cause up-, and down-trends in excess. They decorate the medium- or long-range time charts. And if the antithetical contribution is erased, then almost all the intraday and interday differences fade down.

*Driving:  leaders, media, viruses and worms*

Effects of leaders and media may be considered as driving; since they may have the ability of alligning the society. Furthermore, such drivings may be organized in terms of coalisions of social parties; parallel and coordinated broadcasting of television channels, newspapers, radio stations of the same media group, of a boss or of an holding, etc. On the other hand, antithetical contribution becomes significant under any driving effect, since these individuals oppose the authorithy and slow down her influence within society. When the antithetical contribution is erased, the society becomes (more) easily polarizable under drives. We run the leader effect between 0.0 and 1.0 and vary the number of leaders by 1, 10, 100, and 1000. On the other hand, we run the antithetical percentage from 50 to zero. And we observed that leaders effected the society when their influence parameter is near to 1.0 (more precisely, above 0.8) and when they are crowded (more precisely, when their number is above few hundreds).

When we assume that 20-50% of the society is antithetical, it amounts to 2,000-5,000 people, and when the number of leaders become comparable with that of antithetical people their influence starts to be felt. For (small or) zero antithetical percentages, leader effect starts to be observed at weak influence strengths. Figure 6. a. plots the case, where the antithetical percentage is 50, and leading influence changes from 0.3 to 0.9 in steps of 0.3 for each value of the number of leaders 10, 100, and 1000, consequetively. Figure 6. b. plots the same as Fig.6. a. for an antithetical percentage of zero. It may be claimed that, in case of ideological (and religional, etc.) splittings in society, antithetical people may be considered as resistors to the officially publicated idea. And later, they may become new leaders too, with opposing influence strength to the present ones. Such a society may be considered as having two sets of opposing leaders, or two opposing political parties, etc. We found that, organizing style is the crucial parameter rather than the number and the influencial strength of the leader groups: Any may be greater by a few percent than the other group. Because, maximizing the number of connections necessiates a succesfully made distribution of driving (leading) cites within society. So, in case of close values of numbers and influencial strengths, that leader group with a greater number of effected people per leader wins. This result is important during political elections where two strong parties exist, and percentage of undecided people is high. We also considered instantaneous leadership and media effects with contradiction, i.e., with opposing propositions. Instead of a random distribution of leaders, if they are well organized within a society, they may gain against a mass effect, where the two crucial points are; maximizing the number of connections per leader and establishing stronger bonds.

If, on the other hand, we have a unified group of leaders and if the number of leaders increases in time, then the stituation may be described in terms of viruses and worms as mentioned in previous paragraphs. The condition is that, members of new generation has to be cited with bondings (to leaders) equal to unity.

Figure 7. a. displays how the number of acceptors and rejectors of the idea publicated by viruslike increasing leaders (with a common influence strength of $O_{leader}$ =0.3) and the excess vary in time, for $\delta$=0.01 (number/day). The same plot may be interpretted for the number of buy orders and sell orders, all booked by virus infected computers in market. Figure 7. b. is for a worm, which (with an influence strength of 0.1) moves (infects) with a constant spead of 0.01 (number)/(day). If on the other hand, the rate ($\delta$) is reversed, we have decaying leadership and a dying worm.

It is clear that the strength of leader and media effects is a function of their influence (i.e. magnitude of opinions broadcasted) and that of their number. In some cases their number may vary, and in some other cases their influence, or both. So, one may take into account their number as well as their influence, while studying their driving effect. In the following subsections we will consider some drives which are harmonic and hyperbolic, where the lateral ones can be written as some constant term equals to multiplication of the influence and the number. In this manner one may observe the effect of influence versus the number.

### *Harmonic drives*

A sinusoidal function is known to display a variety of properties over a period: For example it is nearly constant at extrema and nearly linear at middle regions. That is why we tried sinusoidal variation for both the number of leaders, keeping the influence magnitude constant and vice versa, i.e., for the influence magnitude and keeping the number of leaders constant. The complete period of the utilized sinusoidal function covered 8 market weeks in both cases, where 1 market week is composed of 5 market days. So, if $O_{leader}$ stands for the influencial magnitude (opinion of the leaders), $O_{leader}$=0.1 sin(2π(t/8)), where t runs over market weeks, for a constant number of leaders. Here, we observed that the excess roughly followed the periodicity, within a time domain of 16 weeks, i.e. two full sinusoidal periods. We made a similar observation for sinusoidally varying number of leaders with constant $O_{leader}$, i.e., N=ABS(700sin(2π(t/8))), where ABS stands for taking the absolute value, since negative number of leaders does not exist. And, again t runs over (16) market weeks. Furthermore, we took into account two different antithetical percentace of zero and 50, in both of the pronounced sinusoidal driving cases, and we observed again that, high antithetical percentace weakened the present driving effect.

Sinusoidal variations in the number and influence magnitude may exist instantaneously and they may have different angular speeds and phases. These issues may be subject for our future researches. Effect of antithetical percentage on the reponse of the society against the drivig authorities may be another subject for our future researches.

*Hyperbolic drives*

Finally, we considered a case, where both, the leaders' influence and their number are varied with time. We assumed an hyperbolic relation of the form C=N $O_{leader}$, where C is some constant. We run $O_{leader}$ as the independent parameter, which is increased from 0.1 to 1.0 in steps of 0.1 in time (16 market weeks), and C is taken as 10 and 100. We observed that, for a given antithetical percentage, excess decreased with increasing $O_{leader}$, and decreasing N for both of the pronounced values of C. Secondly, increament in antithetical percentage weakened the effect of drives in all of the investigated combinations. It may be noted that, increasing (e.g., doubling) the number of leaders is more effective than increasing (e.g., doubling) their influencial magnitude.

## 4. Conclusion

We do not certainly aim at saying that, any society may be driven according to any albebraic expression. Yet, the algebraically approximated tendencies may involve many clues. The nonlinearity within the response of the society to different algebraic forms may be result of a "social memory". For example, it can be observed that, antithetical people may be playing the role of absorbers (inertia (?)) of the society, against several driving effects.

Many quantities, such as the connectivity and strength of bonds, as well as the percentage of antithetical people in the given society, etc. are needed to known empirically for better utilization of the presented results, else than the assumptions made about the relevant issues.

**Acknowledgement**

The author is thankful to an anonymous referee for his friendly discussions and corrections, and supplying some references.

## FIGURE CAPTIONS

**Figure 1.** Evolution of opinions, where initial opinions (of the society composed by 10,000 citizens) are randomly distributed between -1 and 1. Perpendicular axis is linear at left, and logaritmic at right.

**Figure 2.** Exponential increase of the opinions due to memory effect, where the opinions (with the initial distribution of Fig. 1.) formed during the present interaction tour is summed with the previous one.

**Figure 3. a.** An opinion evolution (with the initial distribution as in Fig. 1.). At the end of each series of 100 tours (day), the given initial distribution is reset. Threshold values of 0.999 and -0.999 are applied.

**Figure 3. b.** Histogram for the opinion distribution of Fig. 3. a. at t=100. Please notify that the evolution saturates early, and that the number of opinion values else than -1, and 1 is very small, yet non-zero.

**Figure 4. a.** An opinion evolution (with the initial distribution as in Fig. 1.). At the end of each series of 100 tours, the given initial distribution is reset. Threshold values of 0.8 and -0.8 are applied.

**Figure 4. b.** Histogram for the opinion distribution of Fig. 4. a. at t=100. Please notify that the evolution saturates early, and that the number of opinion values else than -1, and 1 is very very small, yet non-zero.

**Figure 5. a.** Histograms and tables for the opinion distribution of Fig. 3. a., at t=0.

**Figure 5. b.** Demand, supply and excess, in 5 days (week). For computational parameters see the relevant text and previous figures. Please notify that, demand is ploted with respect to left perpendicular axis, supply is ploted with respect to right perpendicular axis. So, the axes are shifted with respect to each other, but units are the same. Excess (= demand – supply) is also drawn with respect to a shifted perpendicular axis (unshown), with the common unit. Please notice the intraday saturation of demand and supply.

**Figure 6.a.** Leader effect. Antithetical percentage is 50, and leading influence is changed from 0.3 to 0.9 in steps of 0.3 for each value of the number of leaders 10, 100, and 1000, consequetively. Axes as in Fig.5b.

**Figure 6.b.** Antithetical percentage is 0, and the rest is same as in Fig.6.a.

**Figure 7.a.** Virus effect, where the copying rate ($\delta$) is 0.01 (number)/(day). Axes as in Fig.5.b.

**Figure 7. b.** Worm effect, where moving (growing) rate ($\delta$) is 0.01 (number)/(day). Axes as in Fig.5.b.

# FIGURES

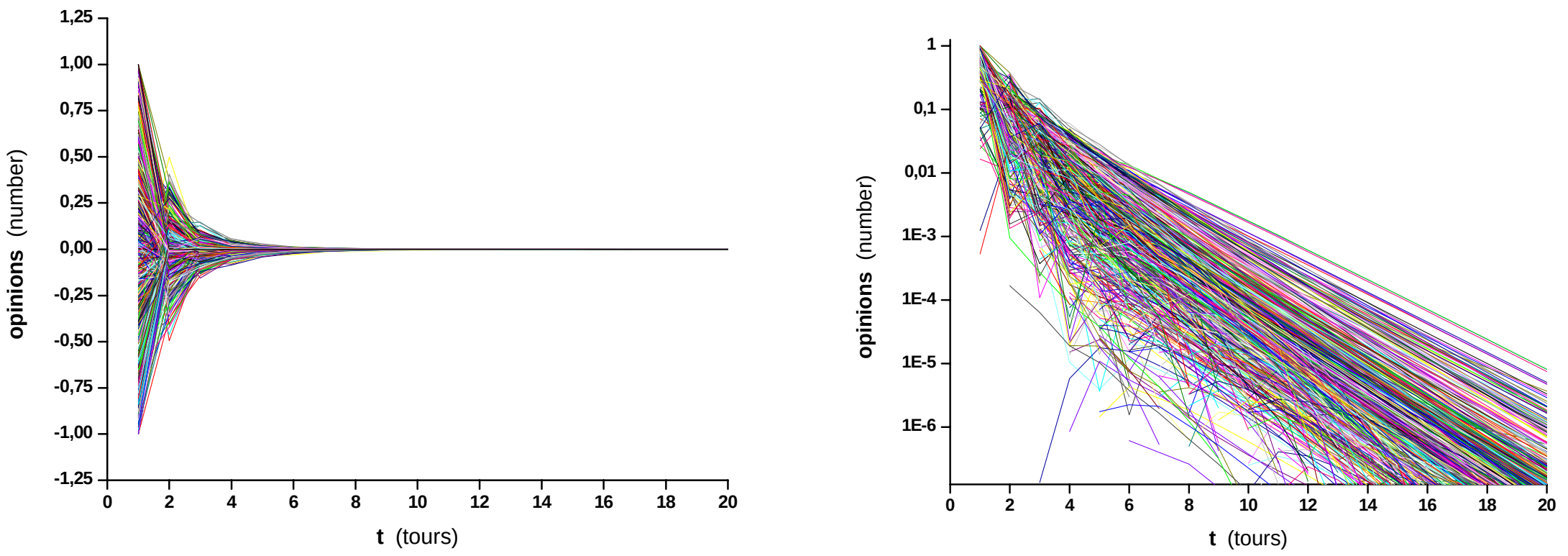

**Figure 1.**

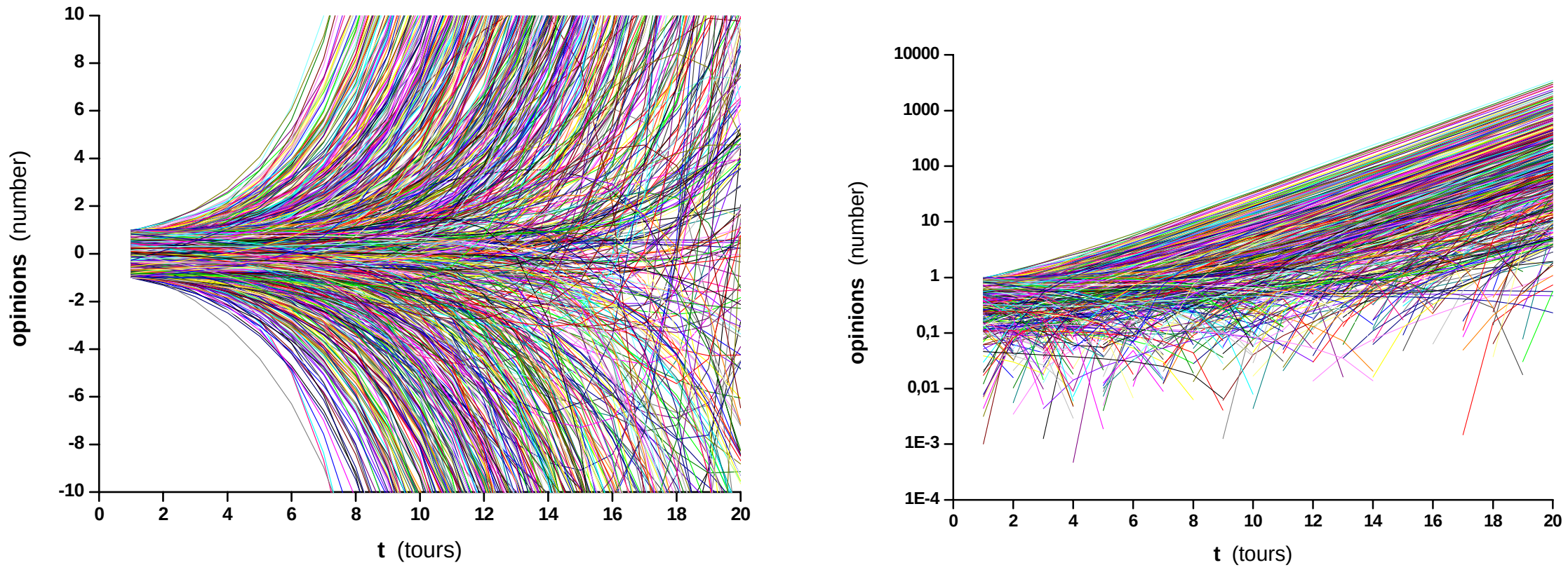

**Figure 2.**

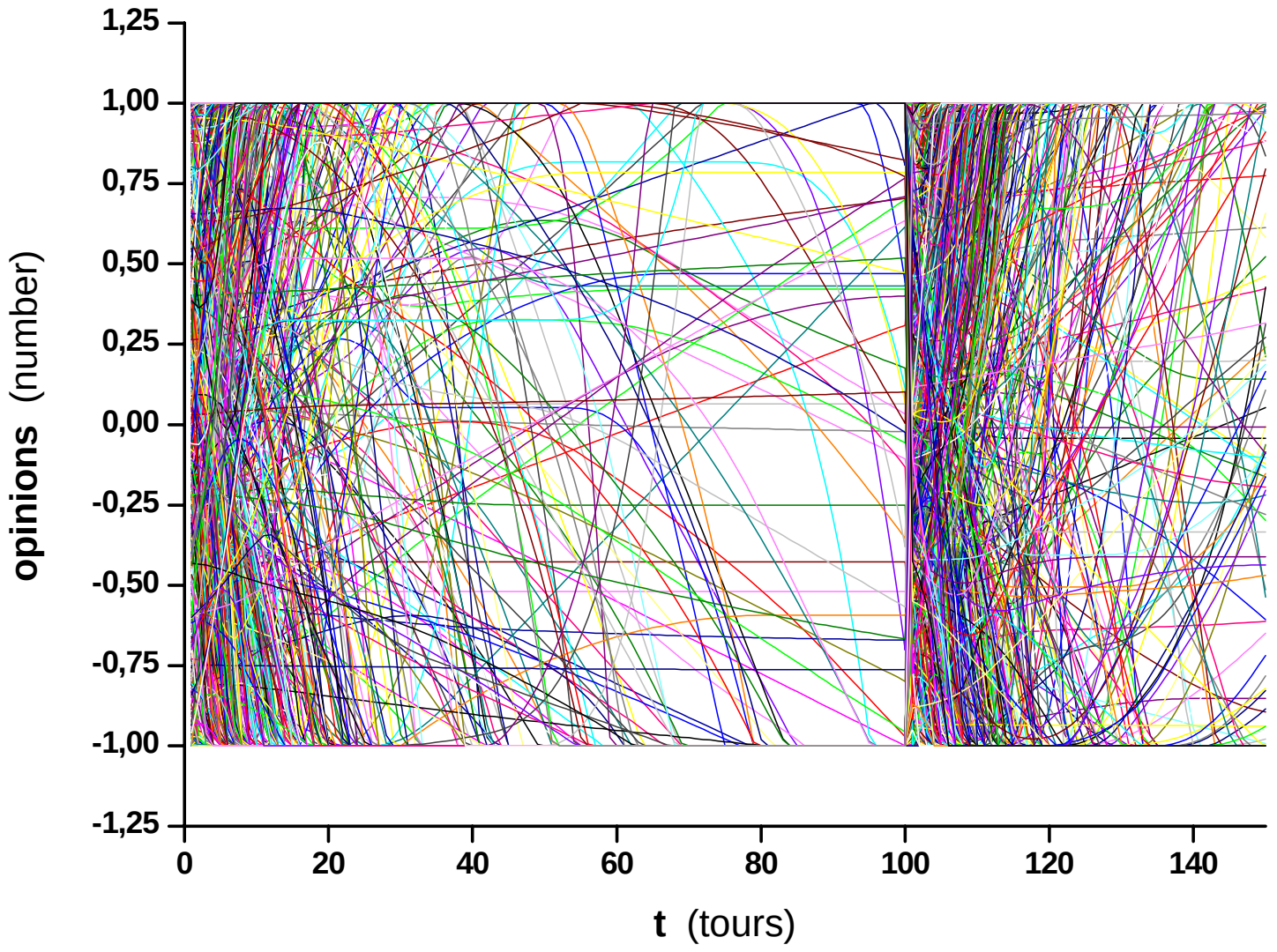

**Figure 3. a.**

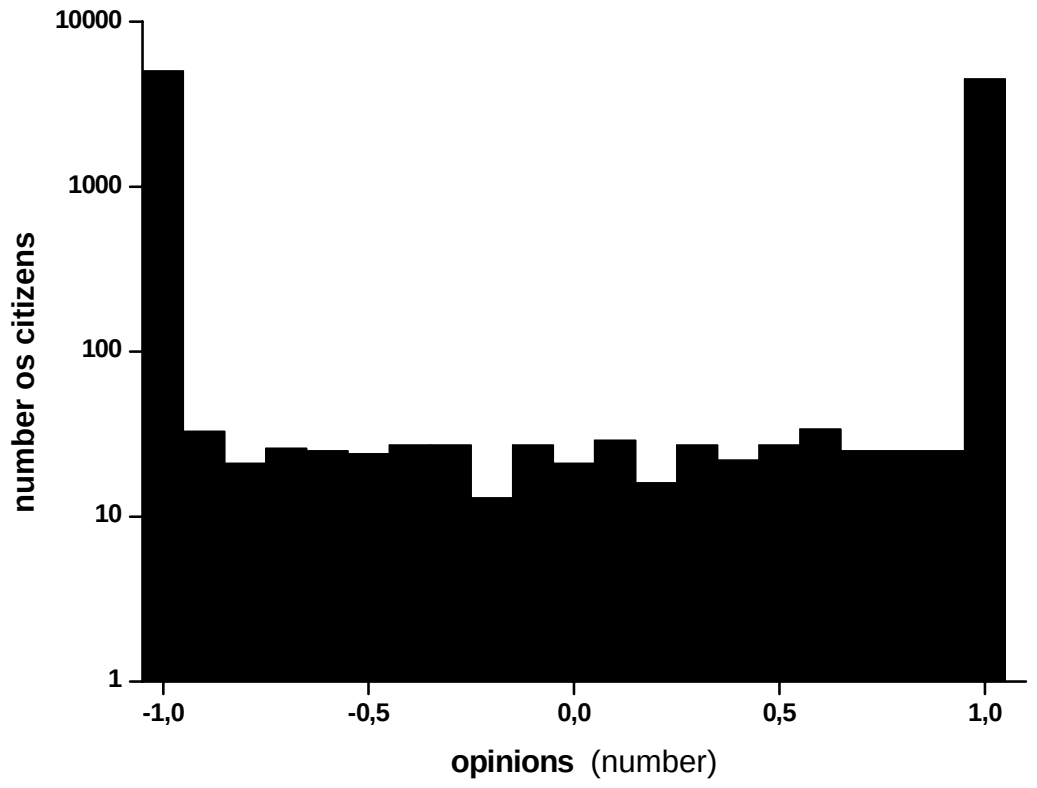

**Figure 3. b.**

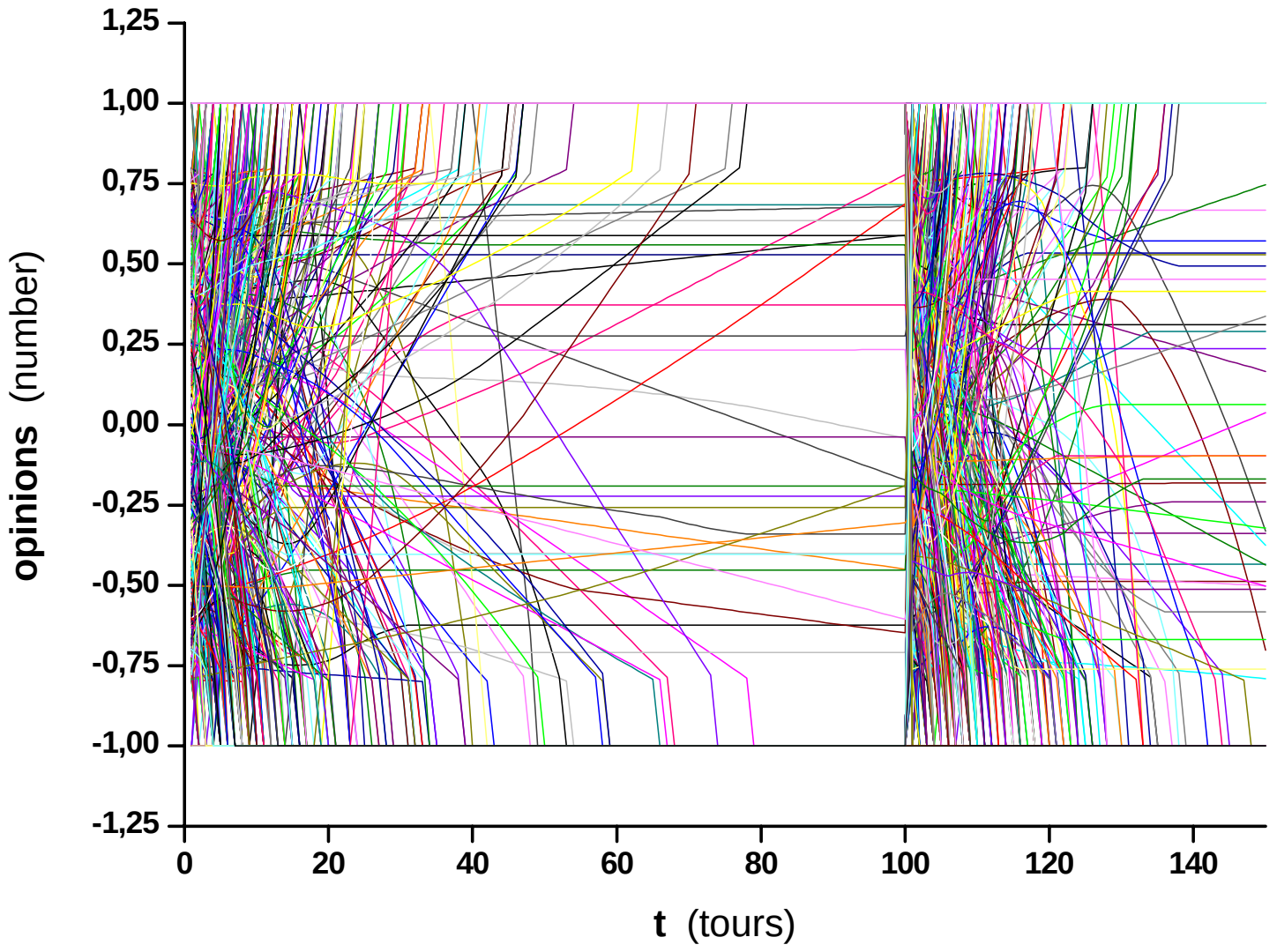

**Figure 4. a.**

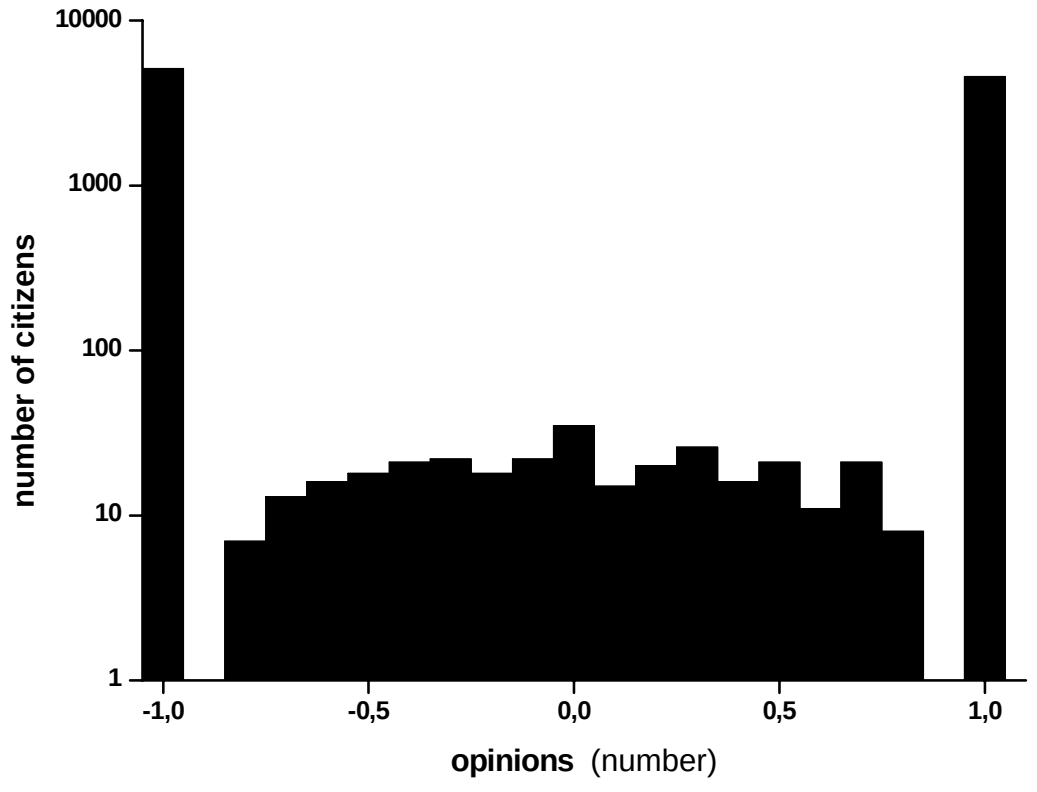

**Figure 4. b.**

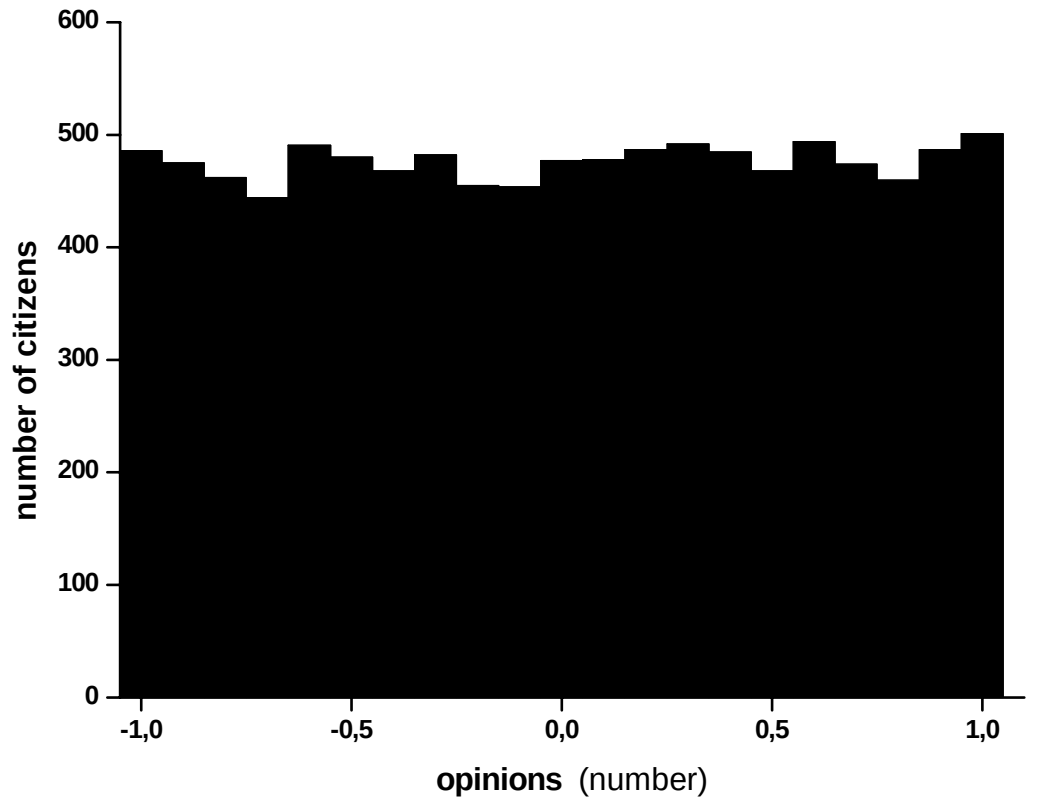

**Figure 5. a.**

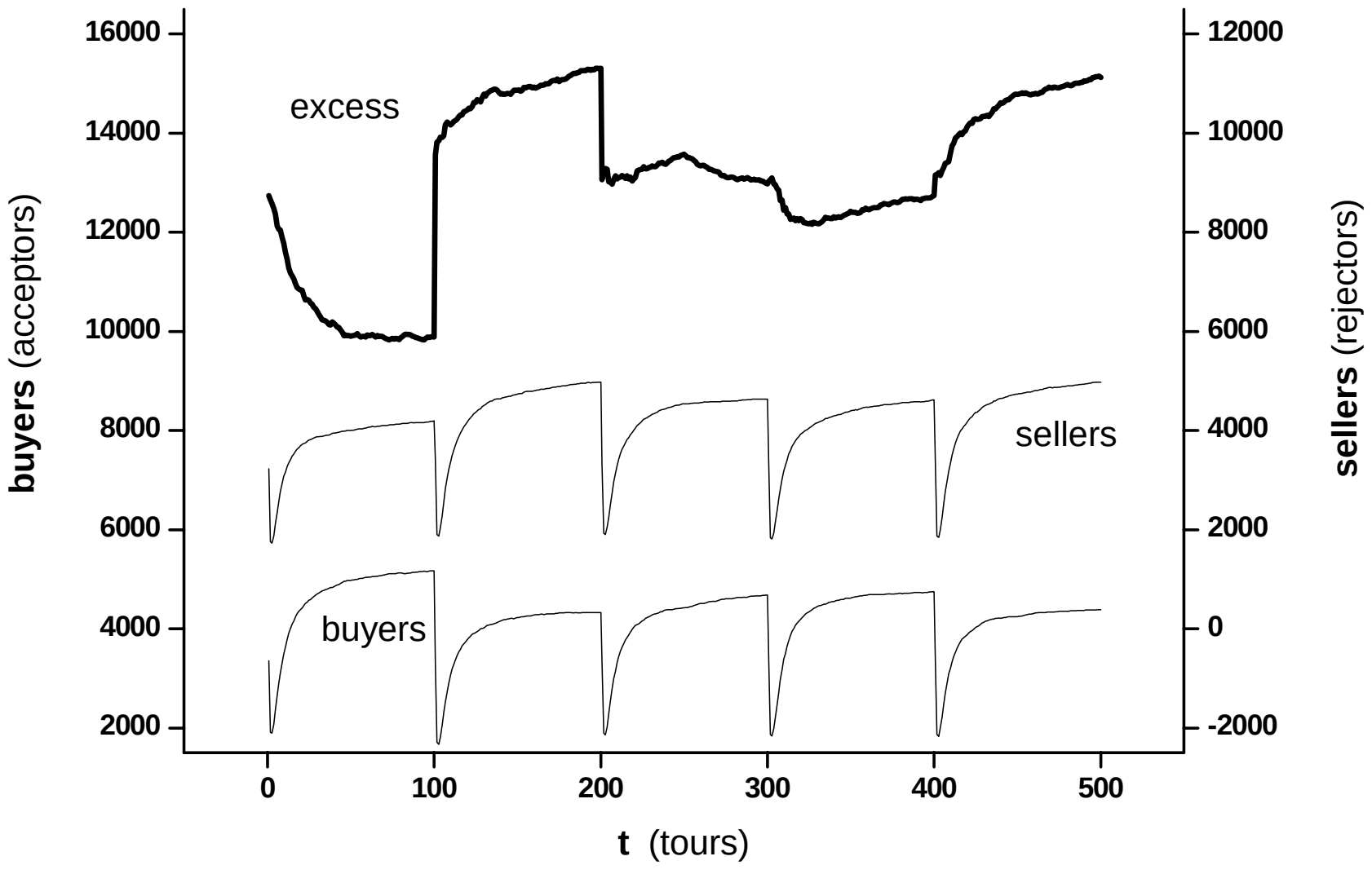

**Figure 5. b.**

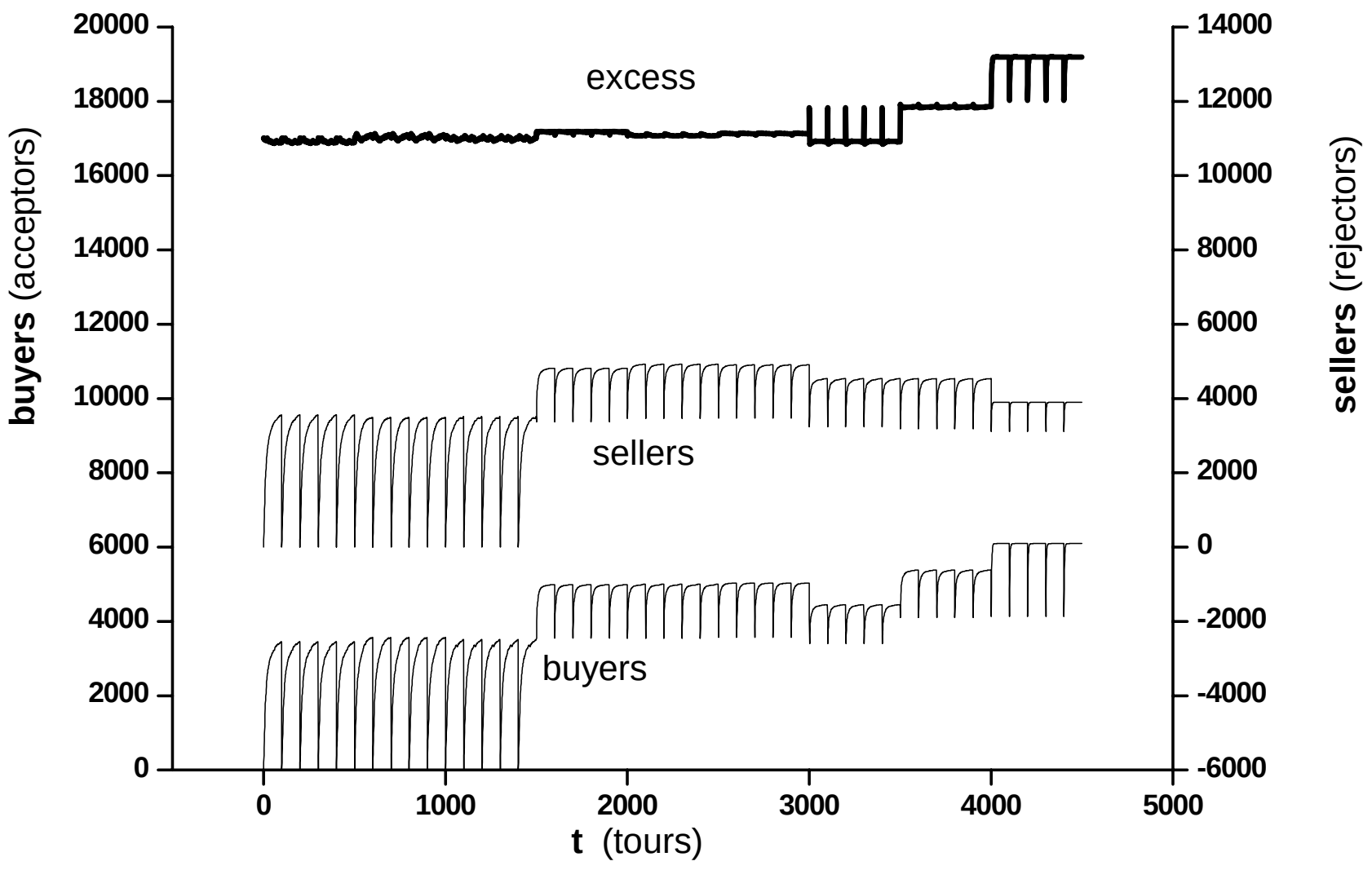

**Figure 6.a.**

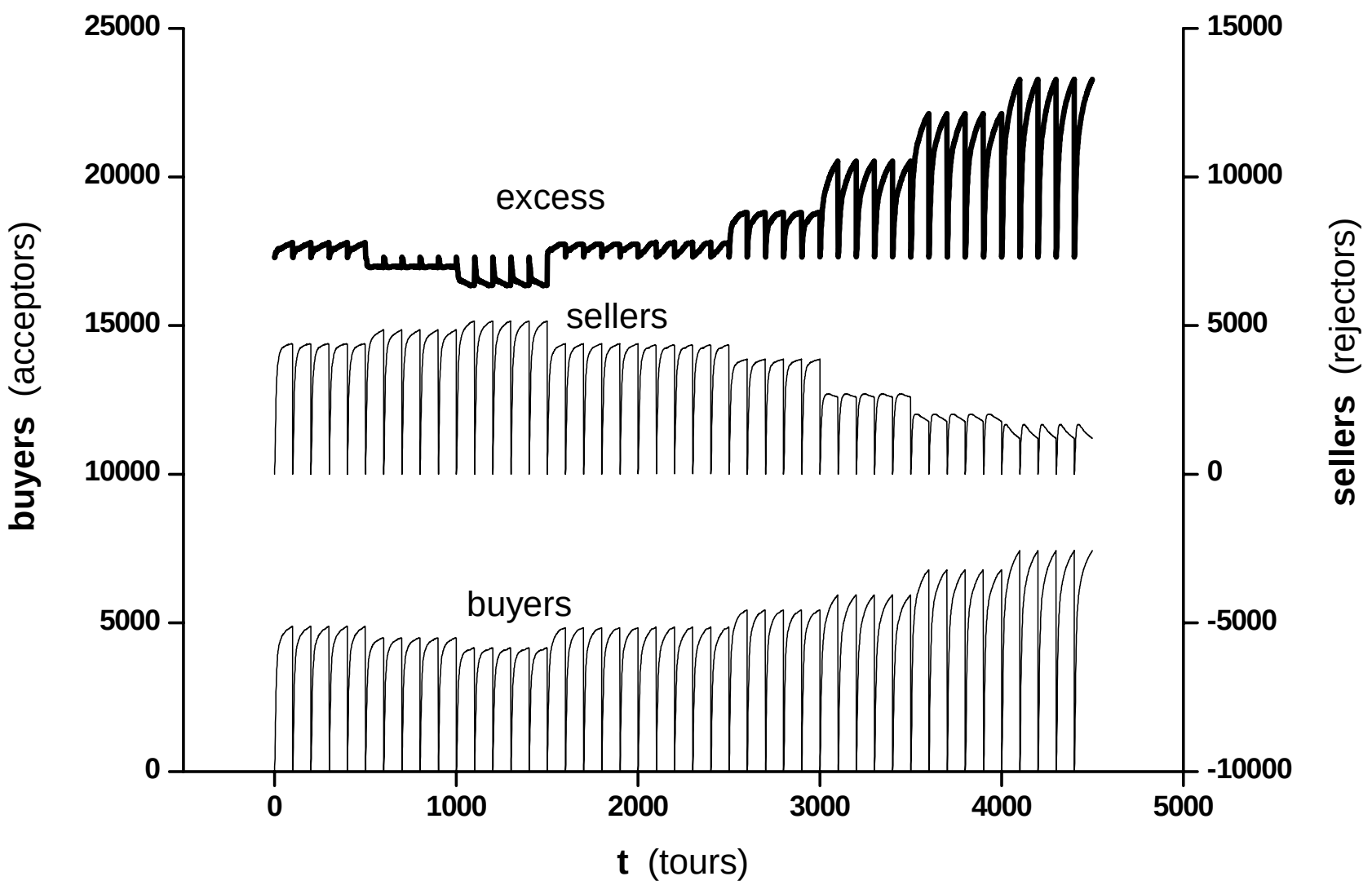

**Figure 6.b.**

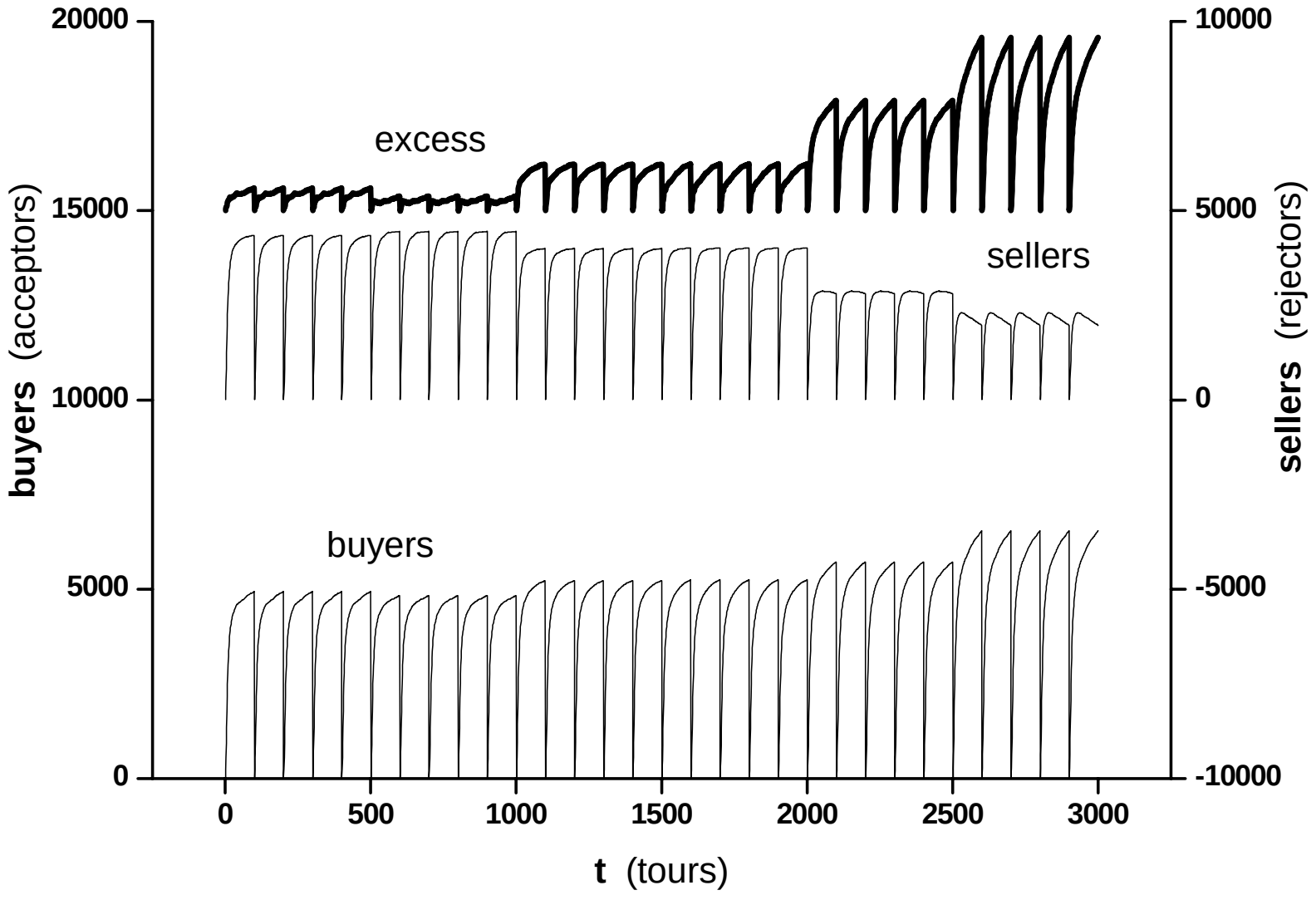

**Figure 7.a.**

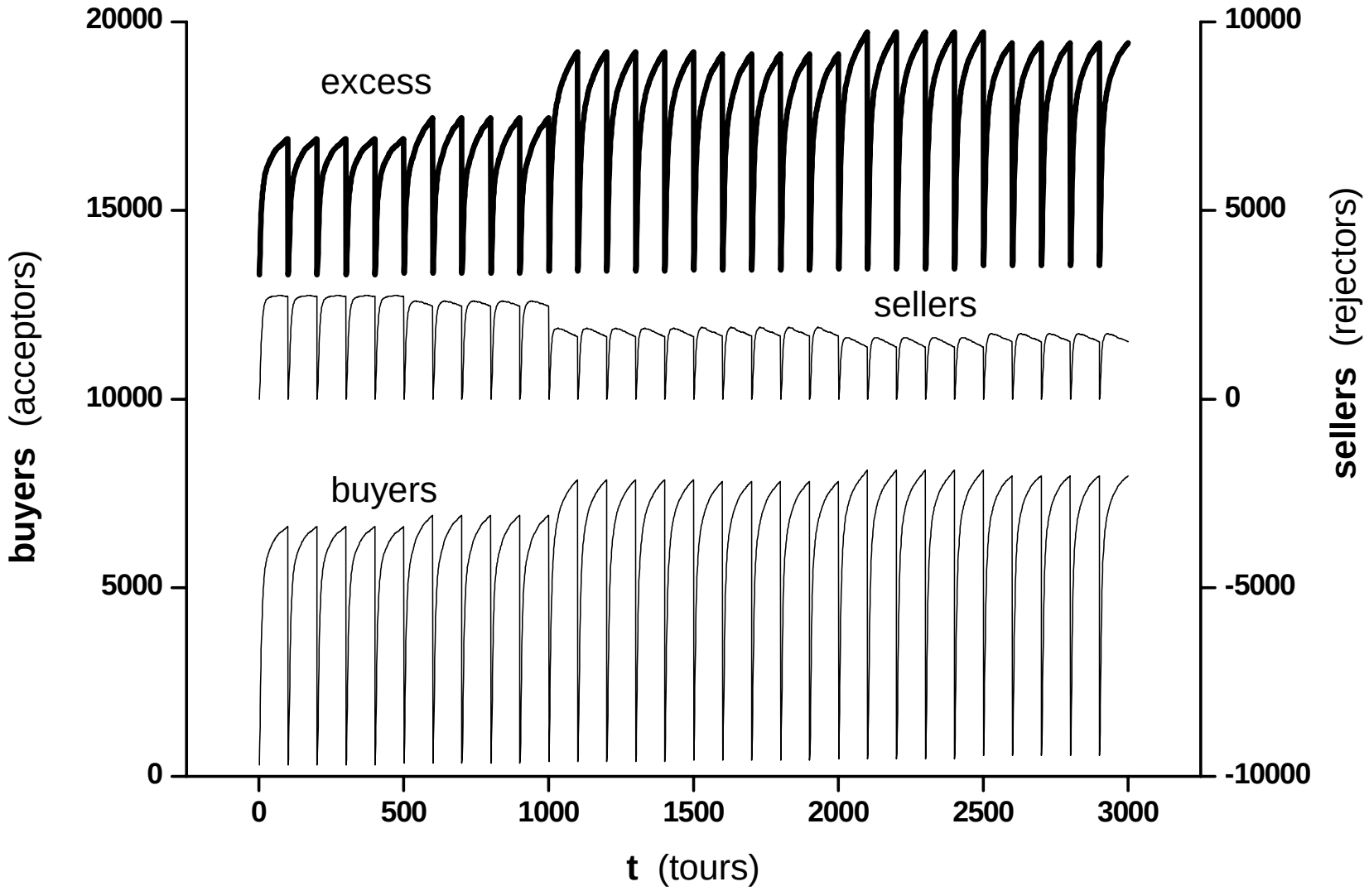

**Figure 7. b.**